\documentclass[preprint,12pt]{elsarticle}

\usepackage{epstopdf} 
\usepackage{color}
\usepackage{epsfig}

\usepackage{pgfplots}
\pgfplotsset{compat=1.14}
\usepackage{pgfplotstable}

\usepackage{subfigure}
\usepackage{amssymb}
\graphicspath{{eps/}}
\usepackage{amsbsy}
\usepackage{amsmath}
\usepackage{amsthm}
\usepackage{color}
\usepackage{booktabs}
\usepackage{graphicx,adjustbox}
\usepackage{algorithm}
\usepackage{algorithmic}
\usepackage{hyperref}
\usepackage{url}
\usepackage{verbatim}
\usepackage{mathrsfs}
\usepackage[a4paper,left=1in,right=1in]{geometry}
\usepackage{setspace}
\usepackage{multirow}
\usepackage[title]{appendix}

\usetikzlibrary{backgrounds,automata}

\allowdisplaybreaks

\journal{TBD}
\begin{document}
\begin{frontmatter}

\title{Agile Earth observation satellite scheduling over 20 years: formulations, methods and future directions}

\address[lab1]{School of Traffic and Transportation Engineering, Central South University, Changsha, China}
\address[lab2]{School of Electronic Engineering and Computer Science, Queen Mary University of London, London, UK}
\address[lab3]{College of Systems Engineering, National University of Defense Technology, Changsha, China}
\address[lab4]{Department of Electrical and Computer Engineering, University of Alberta, Edmonton, Canada}
\author[lab1,lab2]{Xinwei Wang}
\ead{xinwei.wang@qmul.ac.uk}
\author[lab1]{Guohua Wu\corref{cor}}
\ead{guohuawu@csu.edu.cn}
\author[lab3]{Lining Xing}
\ead{xinglining@gmail.com}
\author[lab4]{Witold Pedrycz}
\ead{wpedrycz@ualberta.ca}
\cortext[cor]{Corresponding author}
\begin{abstract}

	\noindent 
	Agile satellites with advanced attitude maneuvering capability are the new generation of Earth observation satellites (EOSs). The continuous improvement in satellite technology and decrease in launch cost have boosted the development of agile EOSs (AEOSs). To efficiently employ the increasing orbiting AEOSs, the AEOS scheduling problem (AEOSSP) aiming to maximize the entire observation profit while satisfying all complex operational constraints, has received much attention over the past 20 years. 
The objectives of this paper are thus to summarize current research on AEOSSP, identify main accomplishments and highlight potential future research directions. To this end, general definitions of AEOSSP with operational constraints are described initially, followed by its three typical variations including different definitions of observation profit, multi-objective function and autonomous model. A detailed literature review from 1997 up to 2019 is then presented in line with four different solution methods, i.e., exact method, heuristic, metaheuristic and machine learning. Finally, we discuss a number of topics worth pursuing in the future.

\end{abstract}


\end{frontmatter}


\section{Introduction}

Earth observation satellites (EOSs) are designed to collect images through their sensors. Due to several advantages such as large-scale observation coverage, EOSs have been broadly applied in the domains of disaster surveillance, environmental monitoring and resource exploration. Meanwhile, the number of orbiting EOSs is continuously increasing and reached 769 in March 2019, contributing to the majority of the orbiting satellites~\cite{2014Nag,UCS2019}. Therefore the EOSs scheduling problem is of great importance and has received much attention over past 20 years~\cite{LemaitreVerfaillie-269,globus2004comparison}.

\begin{figure}[htb]
	\begin{center}
		\includegraphics[width=0.5\textwidth]{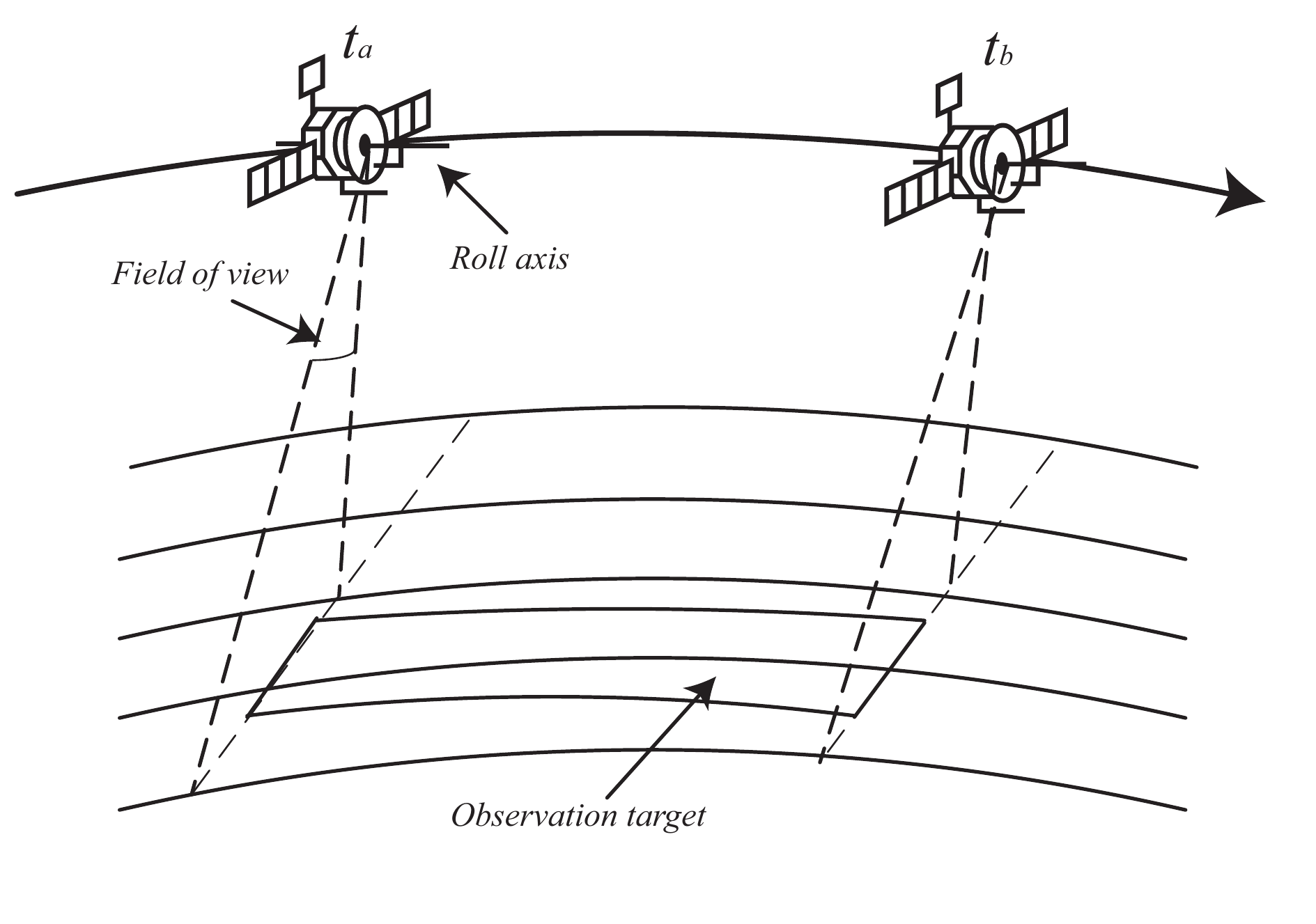}
		\caption{Illustration of the fixed observation interval of a conventional EOS.}\label{figEOS}
	\end{center}
\end{figure}

The agile EOS (AEOS) with stronger attitude adjustment capability along three axes is the new generation of EOS~\cite{LemaitreVerfaillie-269}. Compared to the conventional EOS (CEOS) only maneuverable on the roll axis, AEOS rises a potentially better scheduling efficiency of the observation tasks. As seen in Figure~\ref{figEOS}, the CEOS can only observe the target during a fixed visible time window (VTW) $[t_{a}, t_{b}]$. The interval of the VTW is determined by the satellite and the observation target. Moreover, the AEOS possibly executes two or more observation tasks within a longer VTW, as long as all operational constraints are satisfied. 

\begin{figure}[!h]
	\centering
	\includegraphics[width=4.0in]{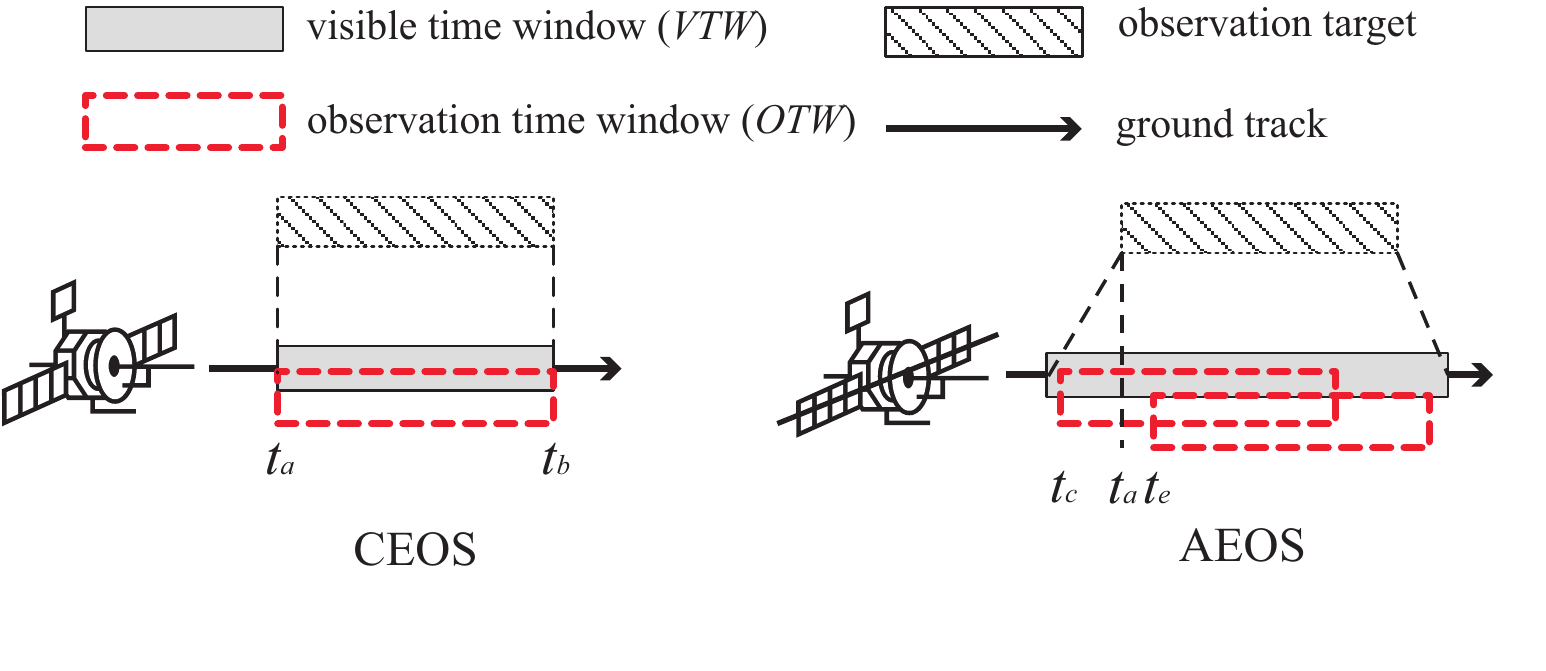}
	\caption{Comparison of the observation capability for CEOS and AEOS.}
	\label{fig:AEOS}
\end{figure}

The typical difference between the observation capacity of CEOS and that of AEOS is illustrated in Figure~\ref{fig:AEOS}. As shown in the left, a CEOS has a specific observation time window (OTW) $[t_{a}, t_{b}]$ to execute the task, which is the same as its VTW. While there is no difference between the VTW and OTW for the CEOS, the VTW for an AEOS is typically longer than the corresponding OTW due to the satellite's ability to look ahead and look back along the pitch axis. Consequently in the right part of Figure \ref{fig:AEOS}, the AEOS can start an observation task for the target at $t_{c}$, or begin observation at $t_{e}$ later than~$t_{a}$. In this way, CEOS scheduling is regarded as a VTW selection problem~\cite{LemaitreVerfaillie-269}, while each VTW contains multiple potential OTWs for an AEOS\@. Although the agile characteristic greatly improves the observation capability of EOS, the complexity of AEOS scheduling problem (AEOSSP) in comparison with CEOS scheduling problem (CEOSSP) also increases dramatically~\cite{liu2017adaptive,wang2018fixed}.

The continuous improvement in satellite technology and decrease in launch cost have boosted the development of AEOSs. One of the most famous AEOS program is Pleiades~\cite{pleiades2019,LemaitreVerfaillie-269}, which was launched in 2003 with the French Space Agency. Pleiades is a two-spacecraft constellation, representing a long-term engagement with the introduction of advanced technologies in Earth observation capabilities. In addition, WorldView~\cite{WorldView} is a next generation commercial imaging task of DigitalGlobe Incorporation, USA. With the addition of its satellite constellation, DigitalGlobe can be capable of collecting around one billion square kilometer of Earth imagery per year. The HJ-1 minisatellite constellation~\cite{HJ-1} is a national program by the National Committee for Disaster Reduction and State Environmental Protection Administration of China, to construct a network of AEOSs. GaoJing/SuperView-1 and -2~\cite{GJ-1} is a commercial constellation of Chinese remote sensing satellites, and it has launched four AEOSs by the end of 2018. The DEIMOS-2 task launched in 2014 by Spain~\cite{Deimos}, aiming to operate an agile small satellite for high-resolution Earth observation applications. To conclude, we further provide some representatives of the orbiting AEOSs and corresponding related works of AEOSSP in Table~\ref{tab:AEOS}.

\begin{table}[htbp]
	\footnotesize
	\centering
	\caption{Representatives of real-world AEOSs.}
	\begin{tabular}{lrllrr}
		\toprule
		Name  & \multicolumn{1}{l}{Launch} & Country & Constellation & \multicolumn{1}{l}{Resolution (m)} & \multicolumn{1}{l}{Related Work} \\
		\midrule
		GeoEye-1 & 2008  & USA   & No    & 0.46-1.84 &  --\\ 
		Pleiades & 2011  & France & Yes   & 1-3   &~\cite{LemaitreVerfaillie-269,GabrelMoulet-258,gabrel2003mathematical,tangpattanakul2012multi,TangpattanakulJozefowiez10}  \\ 
		WorldView & 2007  & USA   & Yes   & 0.3-3.7 &--  \\
		HJ-1  & 2012  & China & Yes   & 30-150 &  \cite{chen2019mixed}\\
		JL-1  & 2015  & China & Yes   & 1-5   &  --\\
		SuperView-1 & 2016  & China & Yes   & 0.5-2 & \cite{liu2017adaptive,peng2019agile,Wang2019,Hanetal18} \\
		DEIMOS-2 & 2014  & Spain & No    & 0.75-20 & -- \\
		KOMPSAT-5  & 2013  & South Korea & Yes   & 1-20  & -- \\
		\bottomrule
	\end{tabular}%
	\label{tab:AEOS}%
\end{table}%

\begin{figure}[!t] 
	\begin{center}
		\pgfplotstableread[col sep=comma]{data_year.csv}\datacsv
		\begin{tikzpicture}
		\begin{axis}[
		xbar,
		y = 0.5 cm, enlarge y limits = {true, abs value = 0.75},
		xmin = 0, enlarge x limits = {upper, value = 0.15},
		xlabel = Number of selected articles,
		ylabel = Publication year,
		xmajorgrids = true,
		ytick = data,
		yticklabels from table = {\datacsv}{a},
		nodes near coords, nodes near coords align=horizontal
		]
		\addplot table [x=b, y=a]
		{\datacsv};
		\end{axis}
		\end{tikzpicture}
		\caption{Overview of number of selected articles in years.}	\label{fig:papers}
	\end{center}
\end{figure}
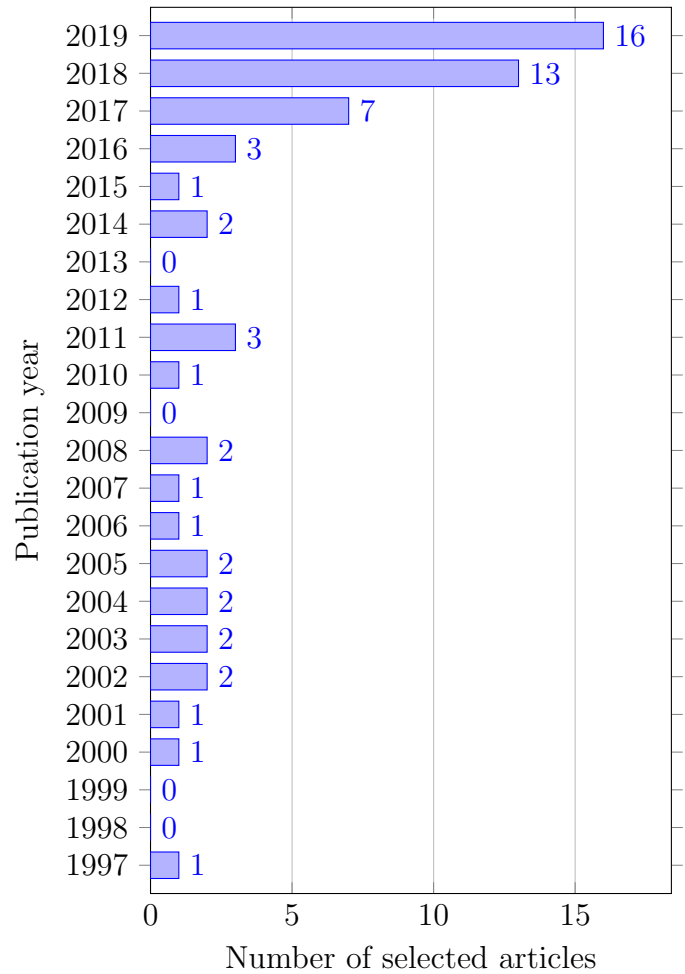

Generally speaking, AEOSSP is to select and schedule satellite observation tasks, aiming to maximize the entire observation profit while satisfying all complex operational constraints. Inspired by the practical demands on effectively and efficiently scheduling AEOSs, the research of AEOSSP has emerged over the past 20 years, especially in the past two years. As seen in Figure~\ref{fig:papers}, we have reviewed 62 articles from 1997 up to 2019 in this work. To the best of our knowledge, Gabrel $et\ al$.~\cite{GabrelMoulet-258} are first to address AEOSSP since 1997, and almost half of the reviewed articles (29 out of 62) have published in the past two years, indicating that the AEOSSP has attracted greater attention than before. However, it is surprising that up until now, there are no survey papers that have addressed the AEOSSP in the open literature. The main reason may be due to the multidisciplinary nature of AEOSSP, making it difficult to conduct a comprehensive literature review. In a practical view, the AEOSSP is a space system problem with aerospace engineering background; while from a theoretic point, the AEOSSP is normally viewed as an combinatorial optimization problem, involving various modeling methods and solution techniques; the AEOSSP can even be characterized as an application of the remote sensing equipment.

To clearly demonstrate the multidisciplinary property of AEOSSP, an overview of the number of selected articles per journal is therefore reported in Table~\ref{tab:papers}. The previous literature has been published in the domains of aerospace engineering, operational research, computer science, remote sensing and multidisciplinary sciences. Researchers in different fields contributed to the same topic, making AEOSSP being of great value both in practice and theory. Moreover, a significant increase of the number of AEOSSP publications, especially in the past two years, also reveals the importance of the AEOSSP.

\begin{table}[htbp]
	\centering
	\footnotesize
	\caption{Overview of the number of selected articles per journal.}
	\begin{tabular}{lr}
		\toprule
		\multicolumn{1}{l}{Journal} & \multicolumn{1}{l}{Number of articles} \\
		\midrule
		AEROSPACE ENGINEERING &  \\
		\midrule
		\textit{Acta Astronautica} & 1 \\
		\textit{Advances in Space Research} & 5 \\
		\textit{Aerospace Science and Technology} & 3 \\
		\textit{International Journal of Aeronautical and Space Sciences} & 1 \\
		\textit{Journal of Aerospace Engineering} & 1 \\
		\textit{Journal of Aerospace Information Systems} & 1 \\
		\textit{Journal of Astronomical Telescopes, Instruments, and Systems} & 1 \\
		\midrule
		& 13 \\
		OPERATIONAL RESEARCH &  \\
		\midrule
		\textit{Annals of Operations Research} & 1 \\
		\textit{Computational Optimization and Applications} & 1 \\
		\textit{Computers \& Industrial Engineering} & 1 \\
		\textit{Computers \& Operations Research} & 3 \\
		\textit{European Journal of Operational Research} & 3 \\
		\textit{Journal of Quality Engineering and Production Optimization} & 1 \\
		\textit{Journal of the Operational Research Society} & 1 \\
		\textit{Naval Research Logistics} & 1 \\
		\textit{4OR} & 1 \\
		\midrule
		& 13 \\
		COMPUTER SCIENCE &  \\
		\midrule
		\textit{Computational Intelligence} & 1 \\
		\textit{Expert Systems with Applications} & 1 \\
		\textit{IEEE Transactions on Evolutionary Computation} & 1 \\
		\textit{Journal of Systems Architecture} & 1 \\
		\textit{Neural Computing and Applications} & 1 \\
		\textit{Swarm and Evolutionary Computation} & 1 \\
		\midrule
		& 6 \\
		REMOTE SENSING &  \\
		\midrule
		\textit{EURASIP Journal on Image and Video Processing} & 1 \\
		\\
		MULTIDISCIPLINARY SCIENCES &  \\
		\midrule
		\textit{Chaos, Solitons \& Fractals} & 1 \\
		\textit{Discrete Dynamics in Nature and Society} & 1 \\
		\textit{IEEE Access} & 4 \\
		\midrule
		& 6 \\
		\midrule
		Overall & 39 \\
		\bottomrule
	\end{tabular}%
	\label{tab:papers}%
\end{table}%

Therefore in this paper, we aim to conduct a comprehensive survey of the AEOSSP and point out possible future research directions. Note that the range satellite scheduling problem (RSSP)~\cite{zufferey2008graph,marinelli2011lagrangian,brown2018heuristic}, which schedules data downloads/communications between satellites and ground control stations, is another satellite scheduling problem similar to AEOSSP. We do not provide detailed review of RSSP, while the potential integrated model of AEOSSP and RSSP will be discussed in Section~\ref{sub:integreateSchedu}. 

The remainder of this document is organized as follows: Section~\ref{sec:definition} presents a general description of the AEOSSP and introduces its typical variations. Section~\ref{sec:review} provides a comprehensive survey with regard to different solution methods, and future directions are pointed out in Section~\ref{sec:future}. Finally, Section~\ref{sec:Concl} concludes this paper.

\section{Agile Earth observation satellite scheduling problem} \label{sec:definition}

In this section, we initially present general definitions of the AEOSSP and model the problem in two different ways to address the tasks transition constraints (which will be defined later). We then extend the basic models by including a series of operational constraints for realistic application, and further introduce three typical variations of the AEOSSP. A summary of the formulation characteristics of AEOSSP is provided in the end.

\subsection{General definitions}\label{sec:def}

To clearly state the general definitions of AEOSSP, several simplifications and assumptions are made as follows.

\begin{itemize}
	\item Multiple AEOSs are required to execute observation tasks on multiple targets during a specified scheduling horizon.
	\item An AEOS can only execute observation task for one target at a given time, and the scheduled imaging tasks cannot be changed.
	\item The candidate observation target is required to be observed for certain times. Typically the maximum desired observation number for one target is one.
	\item Assume that there are enough ground stations and relay satellites in the scenario, therefore the image data download process is not considered in the AEOSSP.
	\item  The tasks transition constraints are defined that sufficient transition time is required between scheduled consecutive observation tasks. These constraints need to be strictly satisfied, otherwise the scheduling scheme results are not feasible. The operational constraints, including onboard energy and memory constraints, are introduced in Section~\ref{sub:cons}.
	\end{itemize}


\begin{table}[htbp]
	\renewcommand\arraystretch{1}
	\footnotesize
	\centering
	\caption{Notation used in the problem formulation.}
	\begin{tabular}{lp{34em}}
		\toprule
		$T$   & Set of observation targets \\
		$i,i^s$   & Observation target index, $i,i^{s} \in T$ \\
		$S$   & Set of agile satellites \\
		$j$   & Satellite index, $j \in S$ \\
		$O_{j}$ & Set of orbits of satellite $j$, $j \in S$ \\
		$k$   & Orbit index, $k \in O_{j}, j \in S$ \\
		$\Theta_{jk}$ & Set of optional observation tasks, $k\in O_{j}, j \in S$ \\
		$p,q$   & Observation task index, $p,q\in \Theta_{jk}$  \\
		$OTW_{jkp}$ & Observation time window for task $p$, $p\in \Theta_{jk}, k\in O_{j}, j\in S$ \\
		$t_{ijk},d_{ijk}$ & Start time and duration of $OTW_{ijk}$, $p\in \Theta_{jk}, k\in O_{j}, j\in S$ \\
		$VTW_{ijk}$ & Visible time window in $k\in O_{j}, j\in S$ for target $i \in T$ \\
		$s_{ijk},e_{ijk}$ & Start and end time of $VTW_{ijk}$, $p\in \Theta_{jk}, k\in O_{j}, j\in S$ \\
	     $\rho_{i}$ &Observation profit of target $i \in T$\\
		$N_{i}$ & Maximum desired observation number for target $i \in T$\\
		$T_{jk}^{sch}$ & Set of scheduled observation targets on orbit~$k\in O_{j}$, $j\in S$\\
		$x_{ijk}$ & Binary decision variable in continues model, $i\in T, k\in O_{j}, j\in S$. $x_{ijk}=1$ denotes that target $i$ is scheduled to be observed within $VTW_{ijk}$, otherwise $x_{ijk}=0$ \\
		$\Delta^{T}(i,i^s)$ & Task transition time from observation task with target $i\in T$ to the immediately subsequent task $i^s\in T$  \\
		$T_{jk}^{sch}$ & Set of scheduled observation targets on orbit~$k\in O_{j}$, $j\in S$\\
		$x_{jkpq}$ & Binary decision variable in time-discrete model, $p,q\in \Theta_{jk}, k\in O_{j}, j\in S$. $x_{jkpq}=1$ denotes that $q$ is the direct successive observation task of $p$, otherwise $x_{jkpq}=0$ \\
		$s_{jk},e_{jk}$ & Dummy start task $s_{jk}$ and dummy end task $e_{jk}$, $k \in O_{j}, j \in S$ \\
		$\Delta_{jkpq}^{T}$ & Task transition time from $p$ to $q$, $p,q\in \Theta_{jk}, k\in O_{j}, j\in S$ \\
		$\Delta_{jkpq}^{V}$ & Attitude transition time from task $p$ to $q$, $p,q\in \Theta_{jk}, k\in O_{j}, j\in S$ \\
		$\Delta_{jkpq}^{S}$ & Attitude stabilization time from task $p$ to $q$, $p,q\in \Theta_{jk}, k\in O_{j}, j\in S$ \\
		$M_{j}^{C},E_{j}^{C}$ & Memory storage capacity and energy capacity during one orbital cycle, $j \in S$  \\
		$M_{j}^{I},E_{j}^{I}$ & Memory and energy consumption for unit time observation of satellite $j$, $j\in S$ \\
		$E_{j}^{M}$ & Energy consumption for unit time of attitude maneuver of satellite $j$, $j\in S$ \\
		$\Theta_{jk}^{i}$ & Set of tasks associated with target $i$ on orbit $k$, $i\in T, k \in O_{j}, j \in S$ \\
		$\rho_{jkp}$ & Observation profit of task~$p$ on orbit~$k \in O_{j}, j\in S$\\
		$d_{jkp}$ & Observation duration of $OTW_{jkp}$,  $p\in \Theta_{jk}, k\in O_{j}, j\in S$ \\
		\bottomrule
	\end{tabular}%
	\label{tab:Notations}%
\end{table}%

The notation used in the definitions of AEOSSP is summarized in Table~\ref{tab:Notations}.
Denote $T$ and $S$ as the set of observation targets and agile satellites, respectively. Let $O_{j}$ be the set of orbits for each satellite $j\in S$. On each orbit $k\in O_{j}$, $\Theta_{jk}$ is defined as the set of candidate observation tasks. Then an tuple $(OTW_{jkp},i,\rho_{i})$ is introduced to represent each candidate observation task $p\in \Theta_{jk}$, where $OTW_{jkp}$ is the OTW of task $p$, $i$ the associated observation target and $\rho_{i}$ the observation profit. The goal of the AEOSSP is to maximize the entire observation profit while satisfying all constraints.

 The required tasks transition time $\Delta^{T}$ is generally composed of satellite attitude maneuver time $\Delta^{V}$ and attitude stabilization time $\Delta^{S}$, and these attributes are all assumed as known parameters~\cite{wertz1978spacecraft}. Note the calculation of transition time depends on the specific OTWs of two tasks, satellite and targets parameters. However, due to the maneuverable flexibility, an AEOS can observe ground target by looking ahead or back along the pitch axis, leading to flexible OTW within its VTW. A detailed analysis in~\cite{liu2017adaptive} demonstrates that even simplified, the tasks transition time is still highly nonlinear for an AEOS\@. To address this issue, the existing AEOSSP formulations in literature are divided into time-continuous and time-discrete models.

\subsubsection{Time-continuous model}

To address the tasks transition constraints, an intuitive idea is to introduce a continuous decision variable $t_{ijk}$ to represent the observation start time within each scheduled VTW with observation target~$i \in T$ on orbit $k \in O_{j}$~\cite{liu2017adaptive,peng2019agile,chen2019mixed}. Besides, decision variables $x_{ijk}$ are defined to check whether $VTW_{ijk}$ is scheduled or not. We define it as the time-continuous model of AEOSSP, which is constructed as 

\begin{align}
\text{max}\quad&  \sum\limits_{i\in{T}} \rho_{i}x_{ijk}  \label{ContinuousObj}\\
\intertext{subject to} &\sum\limits_{j\in{S}} \sum\limits_{k\in{O_{j}}}  x_{ijk} \leq N_{i}  & \forall{i}\in{T} \label{CCon1}  \\
& x_{ijk} (t_{ijk} - s_{ijk}) \geq 0 & \forall k\in{O_{j}},j\in{S},{i}\in{T}  \label{CCon2}\\
& x_{ijk} (e_{ijk} - d_{ijk} - t_{ijk}) \geq 0 & \forall k\in{O_{j}},j\in{S},{i}\in{T}  \label{CCon3}\\
& t_{ijk} + d_{ijk} + \Delta^{T}(i,i^{s}) \leq t_{ijk}^{s} & \forall k\in{O_{j}},j\in{S},{i}\in{T^{sch}_{jk}}  \label{CCon4}\\
& t_{ijk} \in \mathbb{R} & \forall k\in{O_{j}},j\in{S},{i}\in{T}  \label{CCon5}\\
& x_{ijk} \in \{0,1\} & \forall k\in{O_{j}},j\in{S},{i}\in{T}  \label{CCon6}
\end{align}
where objective function~\eqref{ContinuousObj} aims to collect the maximal observation profit. When $\rho_{i}=1$ for each target~$i$, the goal degrades to maximize the number of scheduled observation targets. Constraints~\eqref{CCon1} ensure that each target $i \in T$ is observed at most $N_{i}$ times. Typically the maximum desired observation number for each target is one, i.e., $N_{i}=1$. The observation start time for each scheduled VTW is restricted within its corresponding VTW by constraints sets~\eqref{CCon2} and~\eqref{CCon3}, in which $s_{ijk}$, $e_{ijk}$ are the start and end time of $VTW_{ijk}$, $t_{ijk}$ and $d_{ijk}$ are the start time and duration of OTW within $VTW_{ijk}$. Note that $d_{ijk}$ is given as an input parameter, which is typically less than ten seconds~\cite{LemaitreVerfaillie-269}. The tasks transition constraints are represented in constraints~\eqref{CCon4}, where $\Delta^{T}(i,i^{s})$ is the transition time between two consequent observation tasks and $T_{jk}^{sch}$ represents the set of scheduled observation targets on orbit~$k\in O_{j}$ of satellite $j\in S$.

Clearly this established model is a mixed nonlinear integer problem. Although the multiplication of constraints~\eqref{CCon2} and~\eqref{CCon3} can be resolved by a big-M approach~\cite{chen2019mixed}, the calculation process of the transition time in constraints~\eqref{CCon4} still suffers from highly nonlinear complexity. A feasible solution can be obtained by solving the above model, while a tractable exact method can hardly be designed. The other disadvantage of this time-continuous model is that each VTW is scheduled for at most one observation task, which loses the possibility to have multiple observations for the same target within single VTW. We will further discuss this issue in Section~\ref{subsub:multiobj}.

\subsubsection{Time-discrete model}

The other option to tackle tasks transition constraints is to discretize the VTWs in line with~\cite{GabrelMoulet-258,wang2016scheduling,Wang2019,valicka2019mixed}, where each VTW generates multiple observation tasks for the same target. In this way, each candidate observation task~$(OTW_{jkp},i,\rho_{i})$ has a determined OTW, without introducing the continuous decision variable for observation start time. Thus we only define binary decision variables $x_{jkpq}$ on each orbit $k\in O_{j}$, where both of $p$ and $q$ are indices of the observation tasks. $x_{jkpq}=1$ when $q$ is the direct successive observation task of $p$, and $x_{jkpq}=0$ otherwise. Simultaneously dummy start and end tasks $s_{jk}$ and $e_{jk}$ are added to each orbit.

A preprocessing process is then applied to check the tasks transition constraints in the time-discrete model. For each orbit $k\in O_{j}$, decision variables $x_{jkpq}$ are defined if and only if the observation end time of task $p$ plus the tasks transition time~$\Delta_{jkpq}^{T}$ is not greater than the observation start time of task $q$. The decision variables whose corresponding tasks conflict with each other are removed to satisfy the tasks transition constraints. A time-discrete model of AEOSSP is then developed as 
\begin{align}
\text{max}\quad&  \sum\limits_{j\in{S}} \sum\limits_{k\in{O_{j}}} \sum\limits_{{p\in{\Theta_{jk}^{i}}}}\sum\limits_{q\in \Theta_{jk\cup e}} \rho_{jkp}    x_{jkpq}   \label{DiscreteObj}
\intertext{subject to} &\sum\limits_{j\in{S}} \sum\limits_{k\in{O_{j}}} \sum\limits_{{p\in{\Theta_{jk}^{i}}}}\sum\limits_{q\in \Theta_{jk\cup e}}  x_{jkpq} \leq  N_{i} & \forall{i}\in{T} \label{DCon1}  \\
&\sum\limits_{q\in{\Theta_{jk\cup e}}} x_{jkpq} - \sum\limits_{q\in{\Theta_{jk\cup s}}} x_{jkqp} = \begin{cases}
1, &p=s_{jk}    \\
0, & \forall{p\in{\Theta_{jk}}}\\
-1, &p=e_{jk}
\end{cases}
& \forall k\in{O_{j}},j\in{S}  \label{DCon2}
\end{align}
\begin{align}
&x_{jkpq} \in  \{0,1\}\qquad  \qquad\qquad\qquad\qquad\qquad\qquad\forall k\in{O_{j}},j\in{S}   \label{DCon3}
\end{align}
where $\rho_{jkp}$ is the corresponding target profit of task~$p$ on orbit~$k \in O_{j}$, $\Theta_{jk\cup s}=\Theta_{jk}\bigcup{s_{jk}}$, $\Theta_{jk\cup e}=\Theta_{jk}\bigcup{e_{jk}}$ and $\Theta_{jk}^{i}$ is the set of tasks associated with observation target $i$ on orbit $k\in O_{j}$. 

The objective function~\eqref{DiscreteObj} is to maximize the total observation profit from all targets. Constraints~\eqref{DCon1} indicate that the scheduled observation tasks for target $i\in T$ cannot exceed the required maximum observation number $N_{i}$. The decision variables $x_{jkpq}$ are generated before checking the tasks transition constraints, and the flow constraints are represented by~\eqref{DCon2}. 

This time-discrete model of AEOSSP can be regarded as a specific interval scheduling problem in a parallel machine environment, with each satellite orbit represented as a machine~\cite{Hanetal18}. Notice multiple observations within single VTW is allowed in time-discrete model, as long as the total observation number for target $i$ is not greater than $N_{i}$. However, the discretization of VTWs significantly increases the model scale, and the selection of discretization step needs to be carefully selected (see~\cite{Hanetal18,Wang2019} for more details).

\subsection{Operational constraints} \label{sub:cons}

Practical satellite scheduling involves various complex operational constraints, which can be characterized into two types: temporal constraints and onboard resources limitations~(see \cite{de2006performances,WangReinelt-50,li2019multi,Wang2019} for similar constraints in AEOSSP models).

Temporal constraints actually have been fully considered during the AEOSSP modeling procedure. Since the AEOS operates in a fixed orbit, it only has a limited number of VTWs to the observation target with a definite time duration. The calculation of VTW becomes even more difficult when users raise specific illumination and resolution requirements for the observation tasks. Detailed VTW computation procedures are referred to~\cite{gu2019kriging,han2019visibility,wang2019onboard}. Meanwhile, an AEOS has to download data to a ground station, or use relay satellites to achieve this~\cite{brandel1990nasa,rojanasoonthon2003algorithms}. The data download windows need to be obtained in advance, greatly increasing the complexity of the satellite scheduling. Considering the majority of the existing literature of AEOSSP assumes that there are sufficient resources to ensure data transmission, these constraints are not included in this section. A further discussion of combing data download into AEOSSP will be provided in Section~\ref{sub:integreateSchedu}.

The tasks transition constraints, which have been addressed in Section~\ref{sec:def}, also belong to temporal constraints. Since the AEOSs are utilized as observation resources, a certain amount of
time is required to adjust the imaging field of view to the target from the previous attitude. We need to carefully deal with this kind of time constraints, in case that the scheduled successive observations do not have sufficient transition time.

As for the AEOS onboard resources limitations, the energy and data storage capabilities are critical to the performance of remote sensing system. The onboard energy availability to carry on AEOS is partially supported by a solar panel collecting energy from the Sun. Although the conditions for solar energy collection vary due to environmental variation, the amount of energy collection in one orbit is assumed to be nearly constant~\cite{WangDemeulemeester-4,wang2019robust}. This is the reason why we decompose the observation tasks into different satellite orbits during the AEOSSP modeling, while it is not necessary. We therefore assume that the maximal energy capacity of satellite $j$ is constant and denoted as $E^{C}_j$ in each orbit. The unit-time imaging and maneuvering energy consumption are defined as $E^{I}_j$ and $E^{M}_j$ for satellite $j$, respectively. The number of observation tasks in each orbit has been restricted in order to satisfy onboard energy constraints.

Since the data download process is not considered, we simply assume that satellites can transfer data to the ground station or relay satellite after each orbit, and the data occupation of the scheduled observation tasks in one orbit is limited. Therefore the maximal data capacity in one orbit of satellite $j$ is defined as $M^{C}_j$, and the unit-time imaging data occupation is denoted as $M^{I}_j$. Note the onboard memory constraints can also be limited within the whole scheduling horizon~\cite{liu2017adaptive,Song2018An}.

The extended formulation for AEOSSP including operational constraints can now be constructed for both of the time-continuous and time-discrete AEOSSP models. For instance, the extended time-discrete formulation of AEOSSP is to maximize~\eqref{DiscreteObj} subject to~\eqref{DCon1}--~\eqref{DCon3} and
 \begin{align}
&\sum\limits_{{p\in{M_{jk}}}}\sum\limits_{q\in{\Theta_{jk\cup e}}} x_{jkpq} d_{jkp} M^{I}_{j} \leq{M^{C}_j} & {\forall k\in{O_{j}},j\in{S}} 
\label{NewMemoryCon}\\
&\sum\limits_{{p\in{\Theta_{jk}}}}\sum\limits_{q\in{\Theta_{jk\cup e}}} x_{jkpq}d_{jkp} E^{I}_{j} +\sum\limits_{{p\in{\Theta_{jk}}}}\sum\limits_{{q\in{\Theta_{jk}}}} x_{jkpq}\Delta_{jkpq}^{V} E^{M}_{j} \leq{E^{C}_j} &{\forall k\in{O_{j}},j\in{S}}
\label{NewEnergyCon}
\end{align}
\noindent where $d_{jkp}$ is the observation duration of $OTW_{jkp}$. Similar extensions can also be provided for time-continuous AEOSSP formulation. 

\subsection{Variations}
The AEOSSP has been well studied over the past 20 years, during which the AEOSSP itself has been further extended and modified from the basic model. To provide a clear knowledge of existing AEOSSP formulations, we have reviewed relevant  literature and summarized the three main variations of AEOSSP: different definitions of observation profit, multi-objective function and autonomous models.

\subsubsection{Different definitions of observation profit} 

\begin{figure}[!h]
	\centering
	\includegraphics[width=3in]{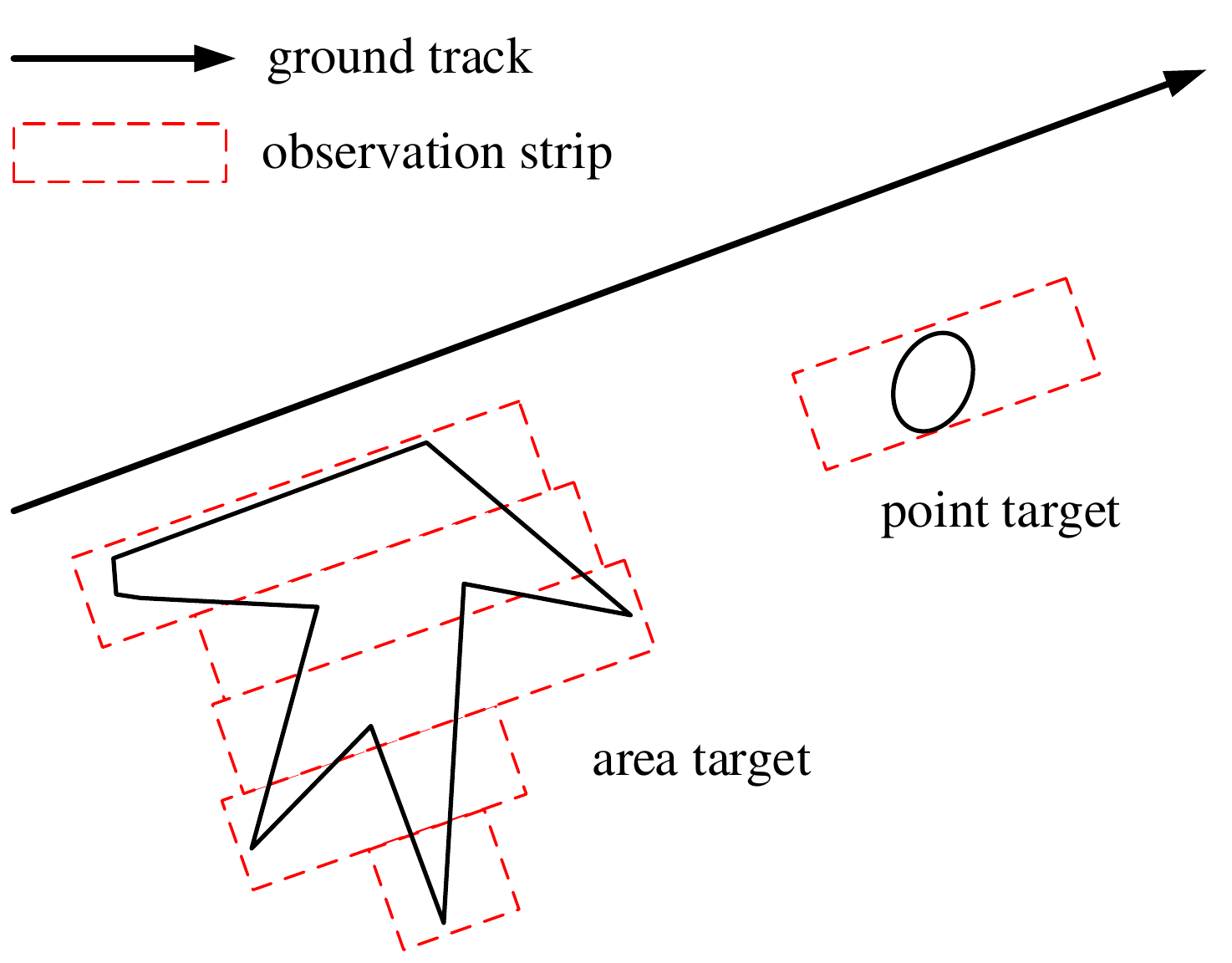}
	\caption{Illustration of observation targets.}
	\label{fig:Areatarget}
\end{figure}

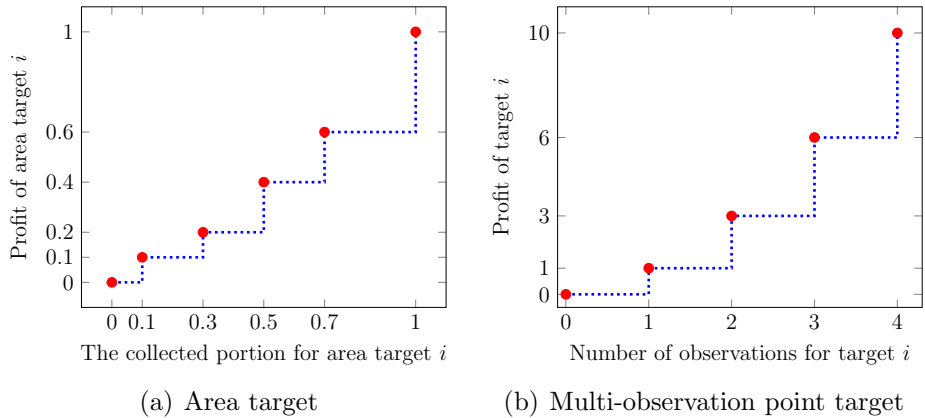
\begin{figure}  [!ht] 
	\centering  
	
	\subfigure[Area target]  
	{  		
		\begin{tikzpicture}[ scale = 0.7]  
		
		\begin{axis}[
		xlabel={The collected portion for area target $i$},ylabel={Profit of area target $i$},
		xmin=-0.1,xmax=1.1,ymin=-0.1,ymax=1.1,
		ytick={0,0.1,0.2,0.4,0.6,1},xtick={0,0.1,0.3,0.5,0.7,1}
		]
		\draw[dotted,blue,line width=0.5 mm] (0,0)--(0.1,0)--(0.1,0.1)--(0.3,0.1)--(0.3,0.2)--(0.5,0.2)--(0.5,0.4)--(0.7,0.4)--(0.7,0.6)--(1,0.6)--(1,1);
		\draw[red,fill=red] (0,0) circle (0.5ex);
		\draw[red,fill=red] (0.1,0.1) circle (0.5ex);
		\draw[red,fill=red] (0.3,0.2) circle (0.5ex);
		\draw[red,fill=red] (0.5,0.4) circle (0.5ex);
		\draw[red,fill=red] (0.7,0.6) circle (0.5ex);
		\draw[red,fill=red] (1,1) circle (0.5ex);
		\end{axis}
		\end{tikzpicture}  	
	}  
	\subfigure[Multi-observation point target]  
	{  
		\begin{tikzpicture}[ scale=.7]  
		\begin{axis}[
		xlabel={Number of observations for target $i$},ylabel={Profit of target $i$},
		xmin=-0.1,xmax=4.3,ymin=-0.5,ymax=11,
		ytick={0,1,3,6,10}
		]
		\draw[dotted,blue,line width=0.5 mm] (0,0)--(1,0)--(1,1)--(2,1)--(2,3)--(3,3)--(3,6)--(4,6)--(4,10);
		\draw[red,fill=red] (0,0) circle (0.5ex);
		\draw[red,fill=red] (1,1) circle (0.5ex);
		\draw[red,fill=red] (2,3) circle (0.5ex);
		\draw[red,fill=red] (3,6) circle (0.5ex);
		\draw[red,fill=red] (4,10) circle (0.5ex);
		\end{axis}		
		\end{tikzpicture}  
	}
	
	\caption{Different definitions of observation profit.}\label{fig:PieceProFun}
\end{figure}

To clearly state the AEOSSP, several simplifications and assumptions are made in Section~\ref{sec:def}. 
However, some targets may be relatively large and the satellite cannot completely observe the target in one image/strip. Following~\cite{LemaitreVerfaillie-269,TangpattanakulJozefowiez10,du2018area}, we define this kind of relatively large-scale target as area target, while the target that can be completely observed by one shot is named point target. As seen in Figure~\ref{fig:Areatarget}, an area target is divided into several smaller strips along the satellite ground track, and then the observation profit of the area target is represented by a nonlinear function, which is a convex function of the acquired surface~\cite{LemaitreVerfaillie-269} (see the left part of Figure~\ref{fig:PieceProFun} as an example). This makes the whole formulation nonlinear, significantly increasing the complexity of the AEOSSP. More related works are refereed to~\cite{bensana1999dealing,cordeau2005maximizing,bunkheila2016new}.

Note that the point observation target is also possibly desired to observed more than once~\cite{Hanetal18,Wang2019} under special circumstances, such as auroral activity surveillance~\citep{GorneyEvans-107} and environmental assessment~\citep{NicholShaker-105,StearnsHamilton-106}. Multiple observations for the same target help to achieve stereo and/or time-series observations, where the observation profit turns to be nonlinear associated with observation count for the same target. As seen in the right part of Figure~\ref{fig:PieceProFun}, this is a similar profit definition in line with the area target~\cite{LemaitreVerfaillie-269}, since we assume that the marginal benefit of an extra observation increases with the observation number, up until a maximum desired observation number for each target $i$, i.e., $N_{i}$ (notice that it is not essential). Although the profit of multiple-observation target is similar to the area target, the linearity of the formulation of AEOSSP can be retained~\cite{Hanetal18}, which does not have a huge impact on the solution method of AEOSSP.

\subsubsection{Multi-objective function} \label{subsub:multiobj}

The original objective function of AEOSSP is to maximize entire observation profit from all scheduled tasks, or to maximize the scheduled observation targets as many as possible. However, multiple goals from different users may be raised according to various applications. As far as we are aware, Tangpattanakul $et\ al$.~\cite{tangpattanakul2012multi,TangpattanakulJozefowiez10} are the first to introduce multiple objectives into AEOSSP. They propose a multi-user AEOSSP, where the users are interested in different observation targets. To address the fairness between different users, they establish a multi-objective model, aiming to maximize the scheduled observation profit, and minimize the profit difference between each pair of users at the same time.

Li $et\ al$.~\cite{li2017preference,li2018multiobjective} propose a modified AEOSSP, in which three objectives including profit, quality and timeliness are considered. Specifically, the first objective is to maximize the entire observation profit; the second one aims to maximize the average quality of the observation tasks, where the observation quality depends on the specific observation time; the last one relies on when the target is observed and how many VTWs it has. Furthermore, Li $et\ al$.~\cite{li2018preference} integrate the quantity (to maximize the entire scheduled observation targets) and balance (minimize the deviation of the overall observation time for each AEOS) rules into the previous research~\cite{li2017preference,li2018multiobjective}, and propose a five-objective AEOSSP. 

Cui $et\ al$.~\cite{cui2018mission} consider the AEOSSP with oversubscribed observation targets, and aim to maximize the number of scheduled observation targets, while minimizing the scheduled tasks transition time. Similarly, Wang $et\ al$.~\cite{wang2019task} consider to maximize the observation targets as well as minimize the energy consumption for observation and tasks transition. More studies are referred to~\cite{yang2018bi,li2019multi,abbas2019agile,eddy2019markov}.

\subsubsection{Autonomous model} \label{subsub:auto}

Most of the currently orbiting EOSs are entirely controlled and scheduled from the ground operation center. However, this does not prevent most of the performed observations tasks to be fruitless because of unforeseen situations~\cite{beaumet2008autonomous}. The autonomous model, which the AEOS itself can take incoming events into consideration in a dynamic manner, is therefore developed and applied in AEOSSP.

By assuming that the AEOS is equipped with a cloud detection instrument and, Beaumet $et\ al$.~\cite{beaumet2008autonomous,beaumet2011feasibility} present a generic architecture to achieve reactive and dynamic decision making. Chu $et\ al$.~\cite{chu2017anytime} develop a bi-satellite autonomous cluster with one for targets detection and the other for target recognition. To schedule the emergency imaging tasks in a relatively short time, Song $et\ al$.~\cite{song2019framework} develop a strategy to distinguish different emergency levels, and implement an autonomous model of AEOSSP. Similarly, Pov{\'e}da $et\ al$.~\cite{poveda2019evolutionary} consider the request of the observation targets may be dynamically changed, and design a simulator to optimize scheduling scheme. Recently, Eddy and Kochenderfer~\cite{eddy2019markov} formulate the autonomous AEOSSP as a semi-Markov decision process, where multi-objective functions and complex operational constraints are considered.

To further speed up the response time of the autonomous system, a long short-term memory-based encoding network and a classification network~\cite{peng2018onboard}, an iteratively linear programming (LP) solver~\cite{she2018onboard}, an edge computing frameworkr~\cite{he2019scheduling}, a hierarchical scheduling method with three stages~\cite{he2019hierarchical}, a Monte Carlo Tree search algorithm~\cite{eddy2019markov} are designed, respectively. 

From the view of autonomous AEOSSP model, decisions are not made before the real-time information is obtained by the autonomous AEOS\@. Such a reactive and adaptive intelligent control system can be of great help to practical satellite scheduling.

\subsection{Summary}
\begin{table}[htbp]
	\centering
	\footnotesize
	\caption{Overview of the AEOSSP formulation characteristic.}
	\begin{tabular}{l r p{0.4\textwidth}<{\centering}}
		\toprule
		Formulation chracteristic & \multicolumn{1}{l}{Number of related articles} & \multicolumn{1}{c}{Related work} \\
		\midrule
		Time-discrete model & 7  & ~\cite{GabrelMoulet-258,gabrel2003mathematical,XuChen-5,wang2016scheduling,xie2019heuristic,Wang2019,valicka2019mixed} \\
			Complex operational constraints & 16    & ~\cite{dilkina2005agile,de2006performances,2007Li,grasset2011building,WangReinelt-50,beaumet2011feasibility,XuChen-5,hosseinabadi2017scheduling,yang2018bi,Song2018An,peng2018onboard,cho2018optimization,valicka2019mixed,Wang2019,li2019multi,eddy2019markov} \\
		Multi objectives & 10   & ~\cite{tangpattanakul2012multi,TangpattanakulJozefowiez10,li2017preference,cui2018mission,yang2018bi,li2018preference,li2019multi,wang2019task,abbas2019agile,eddy2019markov} \\
		Autonomous scheduling & 10     & ~\cite{beaumet2008autonomous,beaumet2011feasibility,chu2017anytime,Song2018An,she2018onboard,peng2018onboard,he2019scheduling,he2019hierarchical,poveda2019evolutionary,eddy2019markov} \\
		\bottomrule
	\end{tabular}%
	\label{tab:formulation}%
\end{table}%

To conclude, we list the AEOSSP formulation characteristics in Table~\ref{tab:formulation}. While most research is on the basis of time-continuous model, the time-discrete AEOSSP model has also been applied in literature due to its linearity.  A number of studies have addressed complex operational constraints, since they are of great importance for practical AEOSSP. Meanwhile, the multiple objectives have received much attention in the domain of AEOSSP recently. 
The variation of AEOSSP with autonomous platform also has a promising opportunity for engineering practice.

\section{Methods}\label{sec:review}

To clearly present the state-of-the-art of AEOSSP, we classify the reviewed literature in line with different solution methods: exact method, heuristic, metaheuristic and machine learning. An overview of the applied methods in AEOSSP is listed in Table~\ref{tab:methods}.

\begin{table}[htbp]
	\centering
	\footnotesize
	\caption{Overview of the applied methods.}
	\begin{tabular}{l r m{6.5cm}}
		\toprule
		Applied method  & \multicolumn{1}{l}{Number of related articles} & \multicolumn{1}{c}{Related work} \\
		\midrule
		Exact method &    9   &  B\&B ~\cite{GabrelMoulet-258,gabrel2003mathematical,chu2017branch,chu2017anytime}, commercial solver~\cite{valicka2019mixed,cho2018optimization,chen2019mixed,kuccuk2019constraint,she2018onboard}\\\\
		Heuristic &  20     &      constructive heuristic~\cite{lemaitre2000manage,verfaillie2001selecting,LemaitreVerfaillie-269,lemaitre2002sharing,tonetti2015fully,de2006performances,WangReinelt-50,XuChen-5,bunkheila2016new,wang2016scheduling,wang2017multi,Wang2019,xie2019heuristic}, time-efficient heuristic for autonomous AEOSSP model~\cite{beaumet2008autonomous,beaumet2011feasibility,nag2018scheduling,Song2018An,he2019scheduling,mok2019heuristic,eddy2019markov}\\	\\	
		Metaheuristic &   31    & 
		GA~\cite{LemaitreVerfaillie-269,2007Li,yuan2014agile,hosseinabadi2017scheduling}, ACO~\cite{du2018area,he2019hierarchical},  EA for multi-objective AEOSSP~\cite{poveda2019evolutionary,tangpattanakul2012multi,TangpattanakulJozefowiez10,li2017preference,li2018multiobjective,li2018preference,zhao2018energy,cui2018mission,yang2018bi,wang2019task,li2019multi,abbas2019agile},
		Tabu search~\cite{habet2003saturated,habet2004solving,cordeau2005maximizing,HabetVasquez52}, local search~\cite{dilkina2005agile,grasset2011building,liu2017adaptive,HE201812,peng2018iterated,peng2019agile,hoskins2017satellite,poveda2019evolutionary}, Russian Dolls~\cite{Benoist2004Upper}\\\\
		Machine learning &   3    & LSTM~\cite{peng2018onboard}, data-driven~\cite{song2019framework}, cooperative neuro-evolution~\cite{du2019data} \\
		\bottomrule
	\end{tabular}%
	\label{tab:methods}%
\end{table}%

\subsection{Exact method}

Currently heuristic scheduling techniques are widely used for the AEOSSP, which typically fail to provide bounds on the quality of the scheduling scheme. To address this issue, one should carefully deal with the nonlinear transition constraints between two consecutive observation tasks. Consequently, two exact methods, branch and bound (B\&B) and mixed integer linear programming (MILP) model with commercial solver have been developed.

Gabrel $et\ al$.~\cite{GabrelMoulet-258,gabrel2003mathematical} are the first to discretize the continuous VTWs and establish a linear model on a directed acyclic graph. Then a B\&B method is designed to solve the problem in a small-size instance. The authors also develop a polynomial-time approximation algorithm for the large-scale AEOSSP. Valicka $et\ al$.~\cite{valicka2019mixed} introduce a novel deterministic MILP model for AEOSSP, with solutions providing optimality guarantees. Note that the VTWs in this work are discretized in the preprocessing stage, and a relatively large discritization time step makes the model tractable. The upper bound obtained by a commercial solver is utilized to calculate the optimality gap. The basic model is then extended to a stochastic MILP model under cloud coverage uncertainty, in which the objective is modified to maximize the entire expectation observation profit. 

Instead of discretizing the VTWs, the following research assumes that the transition time is not time-dependent and only related to the scheduled task sequences. By introducing such assumption, the AEOSSP can be established as a linear model. Chu $et\ al$.~\cite{chu2017branch} consider some operational constraints such as memory and energy constraints are not tight constraints in some cases, thus only remaining the time window and tasks transition constraints in the AEOSSP. To tackle the simplified model, an B\&B algorithm, which can solve the problem to optimality, is proposed. In the framework of the proposed B\&B, a high-quality lower bound is obtained using a look-ahead construction method initially, and three complementary pruning strategies are combined to accelerate the algorithm. However, the proposed exact method can solve up to 55 targets according to the computational results. To realize the observation target recognition over sea, Chu $et\ al$.~\cite{chu2017anytime} consider a bi-satellite cluster with an autonomous low resolution satellite for targets detection, leading an autonomous AEOS with high resolution for target recognition. The main contributions of this work are two folds: establishing a bi-satellite autonomous model for AEOSSP, and provide an anytime B\&B which is based on~\cite{chu2017branch}. A real-life scenario over a 500 km~$\times$ 2000 km sea area within 25 targets is applied to demonstrate the autonomous scheduling for 30 high resolution satellites.

Cho $et\ al$.~\cite{cho2018optimization} consider a constellation of AEOSs with the incorporation of various key operational constraints. Based on the linear assumption for the consecutive tasks transition  constraints, they propose a two-step binary LP formulation and obtain high-quality solutions using a standard MILP solver. The performance of the proposed model exhibits significant improvement of up to 35\% over the results of a general greedy algorithm.

Chen $et\ al$.~\cite{chen2019mixed} address the AEOSSP with multiple AEOSs. The authors define and analyze the conflict indicators of all VTWs between the satellites and observation targets, as well as the feasible time intervals of AEOSs. The problem is then formulated as a MILP model, including constraints from the interdependency between feasible time intervals. The proposed model is directly solved by a commercial solver, obtaining optimal or near optimal solutions.

Considering the complex operational constraints in practice, K{\"u}{\c{c}}{\"u}k $et\ al$.~\cite{kuccuk2019constraint} establish a constraint programming model for single-AEOSSP and solve the problem with a commercial solver. According to the simulation results, the optimal solution is obtained in half hour with up to 55 targets. However, it seems the proposed model cannot tackle instances with larger number of observation targets. 

To satisfy the real-time requirement of onboard autonomous scheduling, She $et\ al$.~\cite{she2018onboard} regard the AEOSSP as a dynamical combinatorial optimization problem, which aims to minimize the slew angle and obtain highest observation profit. A LP solver is applied to solve the established MILP model. Simulations results indicate that the proposed method has better performance compared to genetic algorithm (GA).
\subsection{Heuristic}

Heuristic methods can be used to speed up the process of finding a satisfactory solution. To present a clear introduction of the existing heuristics applied in AEOSSP, we divide them into constructive heuristic and time-efficient heuristic for autonomous model.

\subsubsection{Constructive heuristic}

When complex operational constraints are considered in AEOSSP, constructive heuristics can typically provide high-quality solutions in a short time. Lema\^{i}tre $et\ al$.~\cite{lemaitre2000manage,verfaillie2001selecting,LemaitreVerfaillie-269,lemaitre2002sharing} are the first to comprehensively define the AEOSSP, which is based on the French Pleiades project~\cite{gleyzes2012pleiades}. To solve the proposed highly combinatorial problem, they set out the overall problem and analyze its difficulties initially, and then provide a greedy algorithm, a DP method, a constraint programming approach and a local search algorithm. Note that only the last two methods can tackle all operational constraint in real-world problem.

Following the successful launch and on-going operations of the AEOS DEIMOS-2, Tonetti $et\ al$.~\cite{tonetti2015fully} comprehensively present a capacity analysis and task planning tool for DEIMOS-2. This developed tool can satisfy the need of high-fidelity simulation for practical use. Although this work does not introduce the task planners and operators in details, it demonstrates a real-life application for AEOSSP\@. 
De Florio~\cite{de2006performances} aims to improve the performance of satellite constellations for AEOSSP\@. 
Moreover, complex operational constraints including various time constraints and onboard resources limitations are considered. A simple but effective look-ahead heuristic is presented and it has been tested on different satellite constellation configurations. 

Wang $et\ al$.~\cite{WangReinelt-50} develop a nonlinear model including various operational constraints for AEOSSP\@. This work aims to design a priority-based heuristic with limited backtracking and download-as-needed features, which efficiently produces feasible plans in a very short time. 
Afterwards, Xu $et\ al$.~\cite{XuChen-5} establish a over-constrained model for AEOSSP to maximize the entire observation priorities. Since the observation targets may have priority difference, they develop a constructive algorithm, in which a priority-based sequential procedure contributes to reducing feasibility checking. Moreover, several new priority-based indicators are designed to evaluate the performance of the proposed heuristic. However, the established model is a simplified version of the AEOSSP; each available observation time window only has three fixed observation start time.

Bunkheila $et\ al$.~\cite{bunkheila2016new} provide a new algorithm for area target acquisition by an AEOS\@. The proposed method can be divided into two parts: the area target geometric classification algorithm and a target strips partition method. The directions of strips partition depend on minimizing the number of strips and maximizing the observation time windows. Different AEOS orbits are applied to highlight the efficiency of the proposed algorithm.

Inspired by the complex network theory and time windows discretization method in Grabel 1997, Wang $et\ al$.~\cite{wang2016scheduling,wang2017multi} consider the AEOSSP with oversubscribed observation targets and model it in a directed graph, regarding each node as a candidate observation task. Notice the observation start time for each node is fixed, since the observation time windows have been discretized. In the process of modeling, two factors are defined and applied to the proposed fast approximate scheduling algorithm. Although quantitative insight into the AEOSSP by using the complex networks, this work does not consider practical constraints in real-world scenarios. Therefore Wang $et\ al$.~\cite{Wang2019} further develop a more realistic model with various operational constraints and apply the China's commercial high-resolution AEOS constellation Superview-1 as observation resources. Inherited from~\cite{wang2016scheduling}, a constructive heuristic with a feedback process is designed. The discretization step of observation time windows is also analyzed, and the results indicate the discretization step needs to be carefully selected to strike a balance between the fine model and algorithm efficiency. Xie $et\ al$.~\cite{xie2019heuristic} model the AEOSSP by using the complex network theory, and propose a temporal conflict network-based heuristic, which characterizes the conflicts of the VTWs. This work is similar to~\cite{wang2016scheduling,Wang2019}, which introduce the complex network theory into the AEOSSP for the first time. However, the VTWs are not discretized in the established model, and the observation start time of each target is determined by the proposed heuristic.

\subsubsection{Time-efficient heuristic for autonomous model }

Several heuristics are specifically designed for the autonomous AEOSSP, where real-time scheduling is required. Beaumet $et\ al$.~\cite{beaumet2008autonomous} present a generic architecture for autonomous planning of AEOSSP, including two processes: reactive decision making and deliberative planning. The reactive process allows the system to react to environment stimuli, and the deliberative process uses iterated stochastic greedy algorithm to schedule the observation tasks. Note that such algorithm is naturally anytime while no optimal is guaranteed. Most of the currently active Earth-observing satellites are entirely controlled from the ground. However, this does not prevent most of the performed observations to be fruitless because of the unforeseen presence of clouds. To tackle this, Beaumet $et\ al$.~\cite{beaumet2011feasibility} assume that the AEOS is equipped with a cloud detection instrument and propose a feasibility study of autonomous decision making for AEOSSP\@. A reactive and deliberative architecture used in~\cite{beaumet2008autonomous} is introduced, and the experimental results are conducted on offline and online realistic scenarios based on PLEIADES satellites.

Nag $et\ al$.~\cite{nag2018scheduling} introduce Cubesat constellations for AEOSSP, which is recognized as important solutions to increase number of measurement samples over space and time. As the authors mentioned, the Cubesats now have the ability to slew and acquire images within short time. A modular framework combining orbital mechanics, attitude control and scheduling is proposed. In the scheduling process, a DP with two levels of heuristic is applied based on MILP\@. 

Since a growing number of satellite emergency imaging tasks need to be scheduled in a relatively short time, Song $et\ al$.~\cite{Song2018An} aim to implement a model of autonomous AEOSSP\@. A strategy to distinguish different emergency levels and various quantities is proposed, including three quick-insertion algorithms, e.g., emergency task insertion algorithm, general emergency task insertion algorithm and general emergency task planning \& insertion algorithm. 

To effectively solve the AEOSSP, an edge computing framework is proposed in a flexible manner, where a central node and several edge nodes are included in~\cite{he2019scheduling}. The central node corresponds to the operation center for the whole satellite scheduling system, while each edge node represents an AEOS\@. Based on the density of residual tasks, a constructive heuristic is developed in each edge node, reducing the entire computational time. The proposed framework is promising to deal with autonomous AEOSSP, where the tasks arrive dynamically and real-time response is required.

To address short-horizon single-AEOSSP, Mok $et\ al$.~\cite{mok2019heuristic} propose a simple yet effective method, consisting of sorting, selecting and scheduling. The exact brute force method is utilized to analyze optimality and efficiency of the proposed heuristic. It has potential to adopt the proposed method to autonomous AEOSSP, which requires timely processing.

Recently, Eddy and Kochenderfer~\cite{eddy2019markov} regard the autonomous AEOSSP as a semi-Markov decision process and propose two approximation solution approaches based on forward search and Monte Carlo Tree search. To discover the best possible set of algorithm parameters, a hyperparameter search is designed for different scenarios.

\subsection{Metaheuristic}

Metaheuristic is a higher-level procedure to find, generate, or select a search algorithm that can provide a sufficiently high-quality solution to an optimization problem. According to the reviewed literature, metaheuristics contribute to the majority of the solution methods for AEOSSP\@. Evolutionary algorithms (EAs) and single-point search algorithm are the two major metaheuristics in the research on AEOSSP.

\subsubsection{Evolutionary algorithm}

Due to the simplicity and efficiency, EAs, such as GA~\cite{LemaitreVerfaillie-269,2007Li} and ant colony optimization (ACO)~\cite{du2018area,cui2018mission}, have been extensively applied in the AEOSSP. Particularly, most existing research employs various EAs for multi-objective AEOSSP~\cite{TangpattanakulJozefowiez10,li2018multiobjective}. 

Inspired by successful applications of EAs in scheduling domains, Li $et\ al$.~\cite{2007Li} present a combined GA with simulated annealing (SA) for AESSOP. Simulation results show that the combined GA outperforms GA and SA\@. 
Yuan $et\ al$.~\cite{yuan2014agile} aim to rapidly generate high-quality solutions initially, and then propose GA with several evolutionary strategies to obtain better results. The evolutionary parameters settings, such as population size, crossover and mutation rates, are investigated. 
Different from the existing models, Hosseinabadi $et\ al$.~\cite{hosseinabadi2017scheduling} assume that the observation task preemption is allowed to prevent repetitive area target observations. This introduced preemption policy is evoked when multiple AEOSs have interfering observation strips for the area target. Consequently, a GA based metaheuristic is developed, and the simulation results indicate the proposed preemption policy can bring significant increase of the observation profit.

To deal with the area-target dynamic observation, Du $et\ al$.~\cite{du2018area} view AEOSSP as a travelling salesman problem (TSP) with constraints and propose an end-to-end scheduling approach. The proposed method is initially presented based on a dynamic observation mode and a grid discretization approach. Then the drift angle constraints are introduced to account for the effect of image motion on time delay. An improved ACO is designed to solve the TSP with constraints. 
Given the unexpected environmental changes, such as cloud coverage uncertainty, He $et\ al$.~\cite{he2019hierarchical} establish an hierarchical AEOSSP model, which can satisfy the real-time scheduling requirement. A hierarchical scheduling method with three stages is then proposed based on ACO. First an AEOS is selected for each observation target in the pre-assignment stage, and then the VTW between each target and corresponding AEOS is determined in the rough scheduling stage. In the final stage, the observation start time for each task is selected. Computational tests show the the proposed approach saves calculation time due to avoiding the re-scheduling process. 

As mentioned in Section~\ref{subsub:multiobj}, multi-objective AEOSSP has been received much attention in recent years. Inspired by a French Operational Research Society (ROADEF) challenge competition~\cite{ROADEF}, Tangpattanakul $et\ al$.~\cite{tangpattanakul2012multi} develop a multi-objective optimization model for AEOSSP, where the two objectives, maximizing the entire observation profit and ensuring fairness among satellite users, are considered. A biased random-key GA, including one based on dominance and the other on indicator, is designed and testified on realistic instances from ROADEF benchmark. Tangpattanakul $et\ al$.~\cite{TangpattanakulJozefowiez10} further study the ROADEF challenge, which is a simplified version of the real AEOSSP. Since multi-user fairness is considered in this model, an indicator-based multi-objective local search (IBMOLS) approach is proposed. As an iterated local search method, an approximate Pareto front is generated as an empty set and is updated at the end of each iteration. For each iteration, the nondominated solutions are stored in a archive set, and the fitness value of each individual in the population are computed by a binary indicator. Comparing with the biased random-key GA in~\cite{tangpattanakul2012multi}, the proposed method achieves better performance in less computation time.

Li $et\ al$.~\cite{li2017preference} aim to satisfy different objectives in the AEOSSP, including entire observation profit, total image quality and observing targets as early as possible. The preference-based evolutionary multi-objective optimization methods are then introduced to generate preferable solutions for decision maker, whose preference is considered as a reference point. 
Moreover, Li $et\ al$.~\cite{li2018multiobjective,li2018preference} incorporate decision makers' preferences to the multi-objective AEOSSP, where three objectives are considered: the entire observation profit, the overall quality of the whole scheduling scheme and the timeliness which measures how fast the target is observed. This is continuous work based on~\cite{li2017preference}, and a target region-based multi-objective evolutionary algorithm (TMOEA) is designed to obtain a well-distributed subset of Pareto optimal solutions. 

Zhao $et\ al$.~\cite{zhao2018energy} consider a energy-dependent single-AEOSSP and convert this problem into a dynamic TSP (DTSP) through two mappings from observation angle and energy to city and distance. A time-optimal calculation method is utilized to provide the input of the tasks transition constraints. To solve the DTSP, a hybrid Gauss pseudospectral method and GA is developed, where the former part is to optimize the energy consumption, and the latter one randomly generates feasible solutions. 
Other applications of GAs in multi-objective AEOSSP are referred to~\cite{wang2019task,li2019multi}.

Cui $et\ al$.~\cite{cui2018mission} investigate the emergency AEOSSP with single AEOS over one orbit period. Two objectives, maximizing the entire observation profit and minimizing the total task transition time, are jointly taken into consideration after scalarization. A constraint satisfaction model is then established and a modified ACO with Tabu lists (Tabu-ACO) is designed to solve the problem. 
Yang $et\ al$.~\cite{yang2018bi} consider two objectives, the number of scheduled observation targets and the overall observation images quality for the AEOSSP. The authors develop a constrained optimization model and then design a hybrid coding based differential evolution method for this problem. 

\subsubsection{Single-point search algorithm}
A series of Tabu search algorithms have been developed for AEOSSP\@. Habet and Vasquez~\cite{habet2003saturated,habet2004solving} view the AEOSSP as the vehicle routing problem (VRP) with hard time windows and other constraints. They introduce a Tabu resolution methodology and hybrid it with partial enumerations.  Inspired by the challenge organized by ROADEF, Cordeau and Laporte~\cite{cordeau2005maximizing} develop a Tabu search method for AEOSSP\@. Although the proposed approach is simple, it is easy to be implemented and enables the authors to win the second prize in the ROADEF competition. Habet $et\ al$.~\cite{HabetVasquez52} formulated the AEOSSP as a constrained optimization problem, where the stereoscopic observations are considered. Similar to the work in~\cite{habet2003saturated,habet2004solving}, they combine the Tabu search and partial enumerations. To further improve the solution quality, a secondary problem in minimizing the sum of transition duration between observation tasks is introduced. The upper bounds of AEOSSP are obtained through a dynamic programming (DP) algorithm and utilized to verify the efficiency of the proposed approach.

Other search methods have also been extensively applied in AEOSSP\@. Dilkina and Havens~\cite{dilkina2005agile} regard AEOSSP as an important oversubscribed constraint optimization problem, and consider it with dynamically changing task requirements. Since AEOS allows significant flexibility in observation start time, they design a hybrid scheduling approach which uses permutation-based local search with the propagation of constraints. Grasset-Bourdel $et\ al$.~\cite{grasset2011building} consider to solve AEOSSP with a chronological forward backtrack search which satisfies all physical constraints is designed. An experimental planning tool, called PLANET for PLanner for Agile observatioN satElliTes has been developed by French Space Agency.

Similarly, Liu $et\ al$.~\cite{liu2017adaptive} develop an adaptive large neighborhood search algorithm for single-AEOSSP, including a number of operators to improve the current solution. Specifically, six removal and three insertion operators are designed in this work, and the operators are dynamically selected in each iteration. The time slacks are further introduced to confine the tasks transition constraints. 
He $et\ al$.~\cite{HE201812} further extend the aforementioned model to consider AEOSSP with multiple AEOSs. This is crucial since more AEOSs will come into service in the near future. Therefore an adaptive task assignment mechanism is introduced by defining five assignment operators. There operators are iteratively selected to efficiently assign the tasks to different AEOSs. 
Peng $et\ al$.~\cite{peng2018iterated} introduce a concept of minimal transition time, which only depends on the observation start time of the previous task. Based on this idea, a fast and effective iterated local search (ILS) algorithm is then developed. This work applies the same model established in~\cite{liu2017adaptive} and is tested in the same instances. Comparing to the state-of-the-art algorithm in~\cite{liu2017adaptive}, a significant improvement on average 65\% is achieved. This work has been further extended in~\cite{peng2019agile}, where the observation profit of each target is considered to be time-dependent. A modified bidirectional DP based ILS is integrated to calculate the revised objective function. 


Hoskins $et\ al$.~\cite{hoskins2017satellite} focus on the forest fire monitoring problem, resulting in seeking to optimize a satellite constellation design which minimizes the satellite expected maneuver costs. Then a two-stage stochastic programming model as well as a accelerated L-shaped decomposition approach is presented. 
Similar work can be found in~\cite{zhu2010satellite}.

Considering the satellite operators typically implies dynamically altering requests from different customers, Pov{\'e}da $et\ al$.~\cite{poveda2019evolutionary} investigate AEOSSP under cloud coverage uncertainty, and propose a Hillclimber search algorithm with incremental learning. Using a satellite simulator developed by Airbus, the authors are able to realistically evaluate the proposed methods against the operator baselines. However, the simulation results indicate that the performance of different algorithms varies from the instances. Therefore incorporating domain knowledge can yield improvement for the future research.

By establishing a compact linear model with task interval inequalities and cuts associated with the objective function, Benoist and Rottembourg~\cite{Benoist2004Upper} provide the upper bounds for AEOSSP and further demonstrate them on the benchmark of ROADEF Challenge 2003. A final improved solution is obtained through a Russian Dolls procedure, which is based on the idea of dividing a problem into smaller subproblems that are subsequently solved in an ascending order.

\subsection{Machine learning}

Since machine learning can improve the solution quality by data training, it has promising potential to be applied in combinatorial optimization problems~\cite{nazari2018reinforcement,bello2016neural}. To date, the following three attempts to introduce the machine learning to AEOSSP have been made. Notice that machine learning method cannot tackle operational constraints in AEOSSP yet.

Peng $et\ al$.~\cite{peng2018onboard} point out that most of the existing studies tackle autonomous AEOSSP by search algorithm, while the computational efficiency is usually restricted with the onboard AEOS platform. Instead, they develop a sequential decision-making model and introduce a deep learning-based scheduling method to speed up the autonomous response time. Specifically, they design a long short-term memory (LSTM)-based encoding network to extract the features, and further provide a classification network to generate scheduling scheme. 

Given the potential utilization of the existing data for AEOSSP, a general data-driven framework is proposed~\cite{song2019framework}. The framework is composed of three parts: task assignment, scheduling and task execution. During the core part of the scheduling module, the machine learning method and scheduling algorithm are developed. Notice that the machine learning is based on a data-driven manner with existing data, which provides a better initial solution for the subsequent scheduling algorithm. The authors also mention that the choice of scheduling algorithm is diverse and needs to be analyzed according to specific scenarios.

Considering the complexity of the large-scale AEOSSP with multiple AEOSs, Du $et\ al$.~\cite{du2019data} propose a data-driven parallel scheduling approach, consisting of a probability prediction model, a task assignment strategy and a parallel scheduling manner. Given the historical AEOSSP data, the authors train the prediction model with cooperative neuro-evolution of augmentation typologies. Note that the model training process belongs to preprocessing stage. The assignment strategy is then adopted to decompose multi-AEOSSP into several single-AEOSSPs, which significantly decreases the problem size. The several single-AEOSSPs are finally optimized in a parallel manner. Computational results indicate the the proposed approach presents better overall performance than other state-of-the-art methods.

\section{Future directions}
\label{sec:future}

By comprehensively reviewing the current state-of-the-art of AEOSSP, we can draw some
directions future research in this field. First, there is a strong need for the
development of uncertain model of AEOSSP, combining with existing research
on autonomous model. Second, an integrated AEOSSP model considering data download process is desirable to suitably present practical Earth observation system. Third, the super-AEOS with better observation capability is introduced, and its impact on future AEOSSP research is discussed.

\subsection{Uncertain scheduling}\label{sub:uncertain}


The uncertainty is inevitable for practical scheduling problems~\cite{kochenderfer2015decision}. Particularly, the uncertainty of cloud coverage has a huge impact on the AEOSSP\@. Given optical observation instruments are widely equipped on AEOSs, the images obtained by satellites are easily influenced under cloud coverage~\cite{globus2004comparison}. Ju $et\ al$.~\cite{JU20081196} report that about 35\% of the images acquired by Landsat-7 were useless due to clouds, indicating that the cloud coverage uncertainty should be taken into consideration for AEOSSP.

Several studies of CEOSSP under cloud coverage uncertainty have been conducted. Liao $et\ al$.~\cite{liao2007imaging} establish a stochastic integer programming model to describe the uncertain weather conditions including cloud coverage. Wang $et\ al$.~\cite{WangDemeulemeester-4,wang2019exact} introduce a chance constrained programming model to depict the cloud coverage uncertainty, and further formulate the presence of clouds for observations as stochastic events, establishing a nonlinear model which aims to maximize the expectation value of the scheduled observation profit. Xiao $et\ al$.~\cite{xiao2019two} develop a mixed integer linear programming (MILP) model, in which the influence of the cloud coverage is classified into different levels. Wang $et\ al$.~\cite{wang2019robust} propose a robust model of uncertain CEOSSP on the basis of a budgeted uncertainty set, while preserving the formulation linearity. However, due to high complexity AEOSSP compared to CEOSSP, the models and methods proposed in CEOSSP under cloud coverage uncertainty are difficult to be readily applied to uncertain AEOSSP.

A very limited range of AEOSSP research extending the basic model with cloud coverage uncertainty is found in literature. He $et\ al$.~\cite{he2016cloud} design a cloud avoidance model of AEOSSP, assuming that the real-time cloud information is given in advance. Similarly, Pov{\'e}da $et\ al$.~\cite{poveda2019evolutionary} establish an AEOSSP model, in which the cloud coverage uncertainty is obtained via less accurate historical data. Valicka $et\ al$.~\cite{valicka2019mixed} present the cloud coverage uncertainty via scenarios sampling, and propose several stochastic models to maximize the expected observation profit. Notice all the proposed uncertain models consider to deal with cloud coverage uncertainty in a pure proactive manner~\cite{WangDemeulemeester-4}. Future research could therefore focus on combining scheduling methods under uncertainty with autonomous and reactive platforms. On the other hand, extra observations for the same target are not allowed for the existing scheduling methods. Inspired by the concept of backup observation tasks~\cite{zhu2014fault}, further development of such solution approaches that consider redundant observations for the same target seem highly worthwhile.



\subsection{Integrated scheduling} \label{sub:integreateSchedu}

The image data download process is not considered in the model of AEOSSP given sufficient ground station and relay satellite resources, while the data download arrangement, namely the RSSP, is essential for receiving final observation profit in practice. This is due to that a real-world observation planning procedure of AEOSs consists of two parts: observation scheduling (AEOSSP) and data download planning (RSSP).

 Separately scheduling the imagining and download tasks seems to be a natural choice for this type of real-world complex systems; AEOSSP is already NP-hard and a fully integrated model would turn out to be quite intractable. A considerable part of the literature has looked into AEOSSP without arranging specific data download missions at all~\cite{LemaitreVerfaillie-269,GabrelVanderpooten-61,liu2017adaptive,valicka2019mixed}, indicating that the decomposition by scheduling imaging tasks at first and then arranging the data download is a practical and effective procedure.
 
  Nevertheless, it is always appealing to develop an integrated AEOSSP model. Among the open literature of AEOSSP, two attempts have been made for the integrated AEOSSP\@. Wang $et\ al$.~\cite{WangReinelt-50} have proposed an integrated AEOSSP model, which combines the activities to download image data to a set of ground stations. A constructive heuristic is designed to efficiently produce feasible results, while the algorithm performance is hard to evaluate. Similarly, Cho $et\ al$.~\cite{cho2018optimization} formulate an integrated MILP model, aiming to find a set of observation tasks and corresponding data-download tasks. Due to the linear assumption of tasks transition time, the proposed integrated AEOSSP model actually degrades to the CEOSSP and is therefore can be easily solved by a commercial solver. Meanwhile, several recent research in integrated CEOSSP~\cite{xiao2019two,zhu2019three} has been conducted. However, the integrated AEOSSP still needs to be further investigated.
 
 To conclude, an intractable integrated AEOSSP model requires to be followed with an efficient scheduling method. The existing research in the fields of AEOSSP, RSSP and integrated CEOSSP has provided quantitative insight into the integrated AEOSSP\@. We believe it is promising to present the practical AEOS planning procedure via an integrated AEOSSP model, while an efficient scheduling method needs to be carefully designed.
 
\subsection{Impact of super-AEOS}

In comparison with CEOS, AEOS with stronger attitude maneuvering capability has longer VTWs to ground targets, increasing both of the observation capability and scheduling complexity. As seen in Figure~\ref{fig:Areatarget}, the targets are typically covered by strips along the satellite orbiting directions, so that the imaging equipment on the AEOS can execute observation tasks with fixed attitude.

\begin{figure}[!h]
	\centering
	\includegraphics[width=4in]{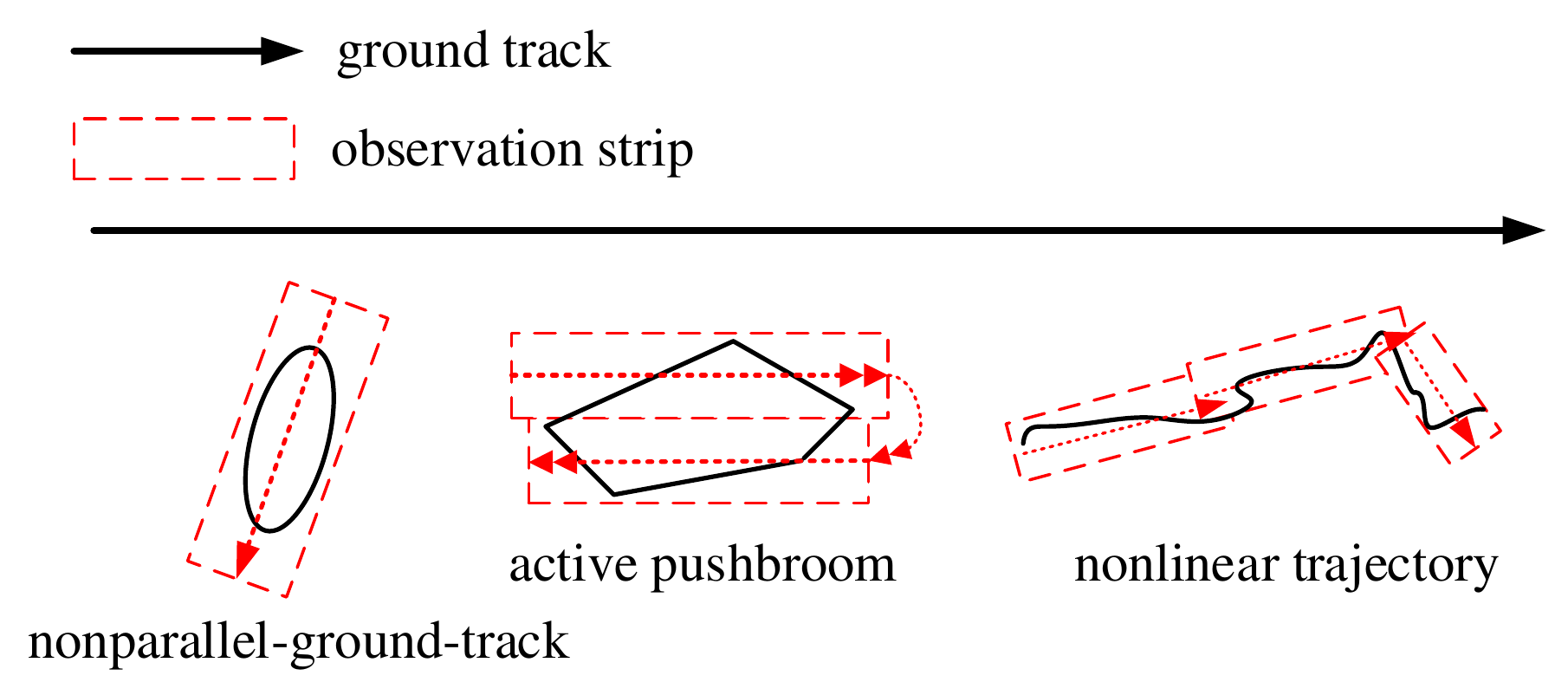}
	\caption{Illustration of super-AEOS observation tasks.}
	\label{fig:SuperAgile}
\end{figure}

To further benefit the agile characteristics of EOSs, the super-AEOS with real-time attitude control system has been investigated recently~\cite{zou2010quaternion,li2013real,zhao2019attitude}. As illustrated in Figure~\ref{fig:SuperAgile}, the developed attitude control system allows super-AEOS to execute the following three types of observation tasks with real-time attitude adjustment: nonparallel-ground-track, active pushbroom and nonlinear trajectory tasks.

When a target is not along the satellite ground track direction, multiple observation strips is required for the AEOS without real-time attitude control system. However, super-AEOS can observe the target in a nonparallel-ground-track manner, achieving more efficient observation task at one time. For the target along with the satellite ground track, an active pushbroom mode of super-AEOS can be utilized to reduce the observation time by continuously adjusting the camera attitude during the imaging procedure. The super-AEOS is also promising to tackle special observation targets such as coastline and borderline, which are presented as nonlinear trajectories.

Therefore the involvement of super-AEOS has huge impact on the AEOSSP, further increasing the scheduling complexity and integrating with the area targets decomposition approach. Although existing research in area observation targets can be referred to~\cite{LemaitreVerfaillie-269,wang2017multi,du2018area}, the active pushbroom mode of super-AEOS with uncertain observation duration clearly has brought new challenges. The definitions of observation profit probably need to be modified in line with the new types of observation tasks. Overall, it is crucial to fully explore the impact of super-AEOS on AEOSSP, and fill the gap in the future research on AEOSSP involving super-AEOS.

\section{Conclusions}
\label{sec:Concl}

The agile Earth observation satellite scheduling problem (AEOSSP) is to select and schedule satellite observation tasks, aiming to maximize the entire observation profit while satisfying all complex operational constraints. The research on AEOSSP has emerged over the past 20 years, especially in the past two years. To summarize current studies and point out future research orientations of AEOSSP, this paper has presented an in-depth literature review of AEOSSP including 62 articles from 1997 to 2019. The general definitions of AEOSSP are initially described, including both time-continuous and time-discrete models. Given various complex operational constraints are involved in practical AEOSSP, the temporal constraints and onboard resources limitations are then included in the basic AEOSSP models, indicating that the problem is NP-hard. This paper also identifies three types of variations of AEOSSP, e.g., different definitions of observation profit, multi-objective function and autonomous model, which have provided promising opportunities for future engineering practice. 

A number of different solution techniques are used in AEOSSP, and we have classified them into four categories: exact method, e.g., B\&B and commercial solver, heuristic, metaheuristic and machine learning.  The heuristic methods including constructive heuristics and time-efficient heuristics for autonomous model have extensively applied in AEOSSP, since they are normally designed for specific optimization problem in practice. Meanwhile, metaheuristics such as EAs and various search algorithms contribute to the majority of solution methods. Most recently, several machine learning methods have been introduced in the domain of AEOSSP, while the operational constraints cannot be well tackled yet.

In terms of future research orientations of AEOSSP, the uncertainty scheduling is inevitable. We believe future research could focus on combining scheduling methods under uncertainty with autonomous platforms, and redundant observations which can be utilized to overcome the task failure deserve further investigation. Meanwhile, it is promising to describe the practical AEOS planning procedure via an integrated AEOSSP model, in which the coupled nature of AEOSSP and RSSP can be better presented. We have also discussed the impact of super-AEOS with stronger observation capability. The super-AEOS will be applied to enhance whole Earth observation system performance, while the scheduling complexity is greatly increased as well. This potential gap is desirable to be filled for real-world super-AEOS applications.

\bibliographystyle{IEEEtran}
\bibliography{SatScheduling}

\end{document}